\documentclass[10pt,twocolumn]{IEEEtran}
\usepackage{multicol}

\usepackage{color}
\usepackage{xcolor}
\usepackage{soul}
\usepackage{amsmath}
\usepackage{graphics}
\usepackage{setspace}
\usepackage{cite}
\usepackage{latexsym}
\usepackage{float}
\usepackage{epsfig}
\usepackage{multirow}
\usepackage{cite,cases,url}
\usepackage{amssymb}
\usepackage{graphicx}
\usepackage{epstopdf}
\usepackage{balance}
\usepackage{subcaption}
\usepackage{enumerate}
\usepackage{array}
\usepackage{algorithmic}
\usepackage{algorithm}
\usepackage{enumitem}
\usepackage{dblfloatfix}
\usepackage[normalem]{ulem}
\useunder{\uline}{\ul}{}

\newcolumntype{L}{>{\centering\arraybackslash}m{5cm}}
\newcolumntype{K}{>{\centering\arraybackslash}m{6cm}}
\newcolumntype{P}{>{\centering\arraybackslash}m{2.3cm}}
\newcolumntype{M}{>{\raggedright\arraybackslash}m{2cm}}
\newcolumntype{N}{>{\raggedright\arraybackslash}m{2.5cm}}

\begin{document}
\title{UAVs with Reconfigurable Intelligent Surfaces: Applications, Challenges, and Opportunities}



\author{
\IEEEauthorblockN{Aly Sabri Abdalla, Talha Faizur Rahman, and Vuk Marojevic}\\ \vspace{0.2cm}
\normalsize\IEEEauthorblockA{Dept. Electrical and Computer Engineering,  Mississippi State University,
Mississippi State, MS\\
}}

\maketitle

\begin{abstract}
A reconfigurable intelligent surface (RIS) is a metamaterial that can be integrated into walls and influence 
the propagation of electromagnetic waves. 
This, typically passive radio frequency (RF) technology is emerging for indoor and outdoor use with the potential of making wireless communications more reliable in increasingly challenging radio environments. 
This paper goes one step further and introduces mobile RIS, specifically RIS carried by unmanned aerial vehicles (UAVs) to support cellular communications networks and services of the future. We elaborate on several use cases, challenges, and future research opportunities for designing and optimizing wireless systems at low cost and with low energy footprint. 
\end{abstract}

\IEEEpeerreviewmaketitle
\begin{IEEEkeywords}
RIS, UAV, advanced wireless, cellular communications. 
\end{IEEEkeywords}

\section{Introduction}
\label{sec:intro}
Next generation networks will embed distributed network intelligence and enable advanced mobile networking and spectrum coexistence among an increasing number of passive and active radio services. 
Unmanned aerial vehicles (UAVs) are to become the killer applications of 5G and the enablers of next generation communications and networking. 
UAVs have unique characteristics, such as higher degrees of freedom regarding positioning and trajectory, low deployment and maintenance costs, and the ability to establish clear line-of-sight (LoS) links with other nodes. 
Therefore, UAVs are considered not only as cellular network users, but also as network 
support nodes for increasing the performance of traditional and new users of mobile networks. 

According to the Third Generation Partnership Project (3GPP), a networked UAV 
can be an aerial base station (ABS), an aerial relay (AR) or an aerial user equipment (AUE). UAVs can be launched to act as an ABS for providing connectivity to UEs, for example when the ground communications infrastructure is damaged 
or unavailable. 
ARs 
are intermediate radio access nodes between the end user and the radio access network (RAN) and send and receive data between a base station and a UE by means of amplify-and-forward AF or decode-and-forward DF. 
AUEs are UAVs that are connected to the cellular network for diverse communications services or command and control signaling. 

Several use cases for UAV assisted wireless networks have been identified by researchers and shown to provide significant performance improvement in terms of throughput, capacity, security, and reliability. 
The research and development activities are continuing to explore diverse innovations to leverage the performance of UAV aided wireless networks and overcome its challenges. 
\begin{figure}[t]
    \centering
    \includegraphics[width=0.48\textwidth]{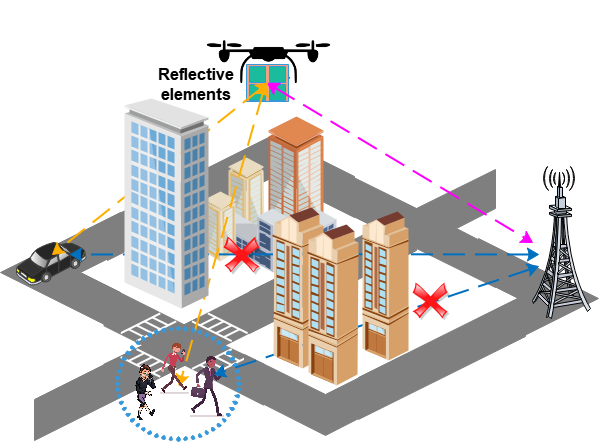}
    \caption{RIS-assisted UAV-aided communications.}
    \label{fig:Figure1}
\end{figure}

The reconfigurable intelligent surface (RIS) is an emerging technology that has the ability to define new wireless transmission and propagation patterns and control the 
communications channel. The RIS is a metasurface that contains 
electronically-controllable and low power consuming 
analog processing elements. The absorption, reflection, refraction, and phase of the 
passive reflecting elements can be adjusted in real time to steer the incident electromagnetic signals in the desired direction. 
The phase and amplitude of the reflected signal 
maximizes the effective channel gain. 
A RIS can be deployed at different locations and attached to surfaces such as buildings, vehicles, and indoor walls with minimal cost and effort. 
An additional advantage 
is its compatibility with current radio technology. 
It 
supports advanced wireless waveforms and both 
full-duplex and half-duplex communications over wide bandwidths and frequency ranges, and many design options exist for RISs. 


\textcolor{black}{
The RIS-assisted UAV communications and networking framework has enormous potential for advancing wireless communications and tackling
the increasing complexity of the wireless channel, especially at higher frequencies as shown in Fig.~\ref{fig:Figure1}. 
RIS-assisted UAV systems can improve the channel quality in urban environments as has been shown in \cite{li2020reconfigurable}. The authors derive a closed-form solution for the phase shifter as function of the UAV trajectory and show how it significantly increases the achievable rates over time. Similarly, \cite{ma2020enhancing} demonstrates how RIS-assisted UAV communications can help improving cellular services. 
The RIS-UAV framework has a considerable reduced power consumption over active ARs, which can be further improved if a joint optimization of UAV trajectory and resource allocation is considered \cite{cai2020resource}.}
\textcolor{black}{ 
Other recently published works 
are specific to solving a particular problem when deploying RIS mainly in a geo-stationary setting. In order to understand the full potential of the RIS-UAV framework, a broader aspect is considered here.}
To the best of our knowledge this work is the first to discuss the breadth of use cases and applications of integrating RIS into UAV nodes in the context of next generation cellular networks. 

The remainder of this paper is organized as follows. Section II describes emerging wireless communications and networking use cases for RISs carried by networked UAVs and their
impact. 
Section III captures the key research challenges and directions to explore, whereas Section IV discusses next generation wireless technology and the role of RISs. 
Section IV derives the conclusions.

\section{
Applications of RIS Assisted UAV Communications }
\label{sec:APPLICATIONS}


There are multiple communications and networking use case where the integration of RISs and UAVs can be beneficial. Here we discuss the impact that this framework can have on 
coverage, massive multiple access, physical layer security (PLS), and simultaneous wireless information and power transfer (SWIPT).


\subsection{RIS-assisted UAV Communications for Extended Coverage}
The ABS and AR solutions are seen as key enablers for providing dynamic and adaptive coverage in future cellular network deployments. Such active aerial communications however incur a significant energy overhead. The RIS is the passive alternative. 

\begin{figure}[t]
    \centering
    \includegraphics[width=0.48\textwidth]{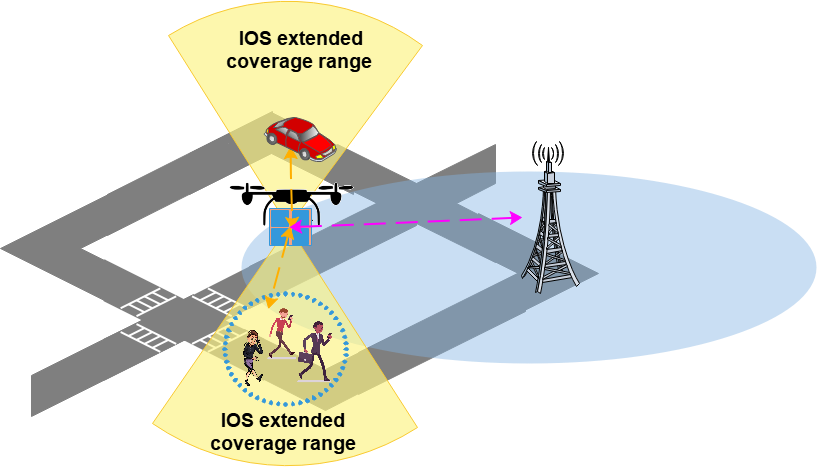}
    \caption{RIS-UAV for extended coverage.}
    \label{fig:Figure2}
\end{figure}

A special case of RIS, which is known as intelligent omni-surface (IOS), has antenna elements on both sides of the meta-surface and is able to reflect incident signals coming from opposite directions. \textcolor{black}{An unobstructed IOS
reflects the incident signals on both sides 
of the 
sheet and can cover dead zones and provide massive 360 degree coverage and higher spectral efficiency.} 
A UAV provides this capability by carrying the IOS underneath it and being able to fly at suitable heights to provide reflective RF surfaces where needed. 
The greatest strength point of the IOS is its ability to \textcolor{black}{control the direction of the departure signal from the RIS to the potential receivers without any blind spots by adjusting the phase shift vectors on either side. The deployment of the RIS-assisted UAV technology at the edge of a base station coverage area and extend the reach of the incident signal SNR. 
The optimization of the UAV trajectory and phase shift vectors 
for reaching the desired UE locations, whether static or mobile, effectively extends the original cell coverage in the desired direction as illustrated in Fig.~\ref{fig:Figure2}.}
This special characteristic of the IOS can be considered as a key enabler to extend the downlink communications coverage of ground base stations 
to serve 
dispersed ground UEs, or clusters of UEs. 

\subsection{RIS-assisted UAV Communications for Increased Capacity}
An AR 
can be positioned to increase the achieved rate on both uplink and downlink between ground BSs and UEs. 
Different optimizations can be deployed for the UAV position, trajectory, transmit power, and orientation. 

The RIS-UAV 
is an alternative way to improve the throughput and capacity. Generally, the RIS functions in full-duplex (FD) mode which improves the spectral efficiency over the half-duplex (HD) mode that is commonly used for ARs. Moreover, the passive nature of the RIS makes its relaying strategy overcome any antenna noise amplification and self-interference, which translates in less computation  and lower power consumption than for active FD relays. 
Interference cancellation can be done by controlling the phase shifts of some of the RIS elements to inverse the interference signal and remove or weaken it, under the assumption of known channel state information (CSI) at the RIS controller. In addition, \textcolor{black}{the RIS can contribute alongside with the UAV to introduce rich scattering of LoS links for several ground UEs by optimizing the phase shifts of the antenna elements.}
As a result of the characteristics of the RIS and, in addition, the LoS capabilities that are enabled by the UAV, the spectral efficiency will exceed the current solutions for increasing capacity. 
Scalability can be achieved with multiple UAVs, combined with static RIS where available.   \textcolor{black}{It is worth mentioning here that LoS links with good channel quality (high SNR) can further enhance the spectral efficiency with the help of aerial-RIS by incorporating spatial multiplexing and/or multi-user MIMO.}

\subsection{RIS-assisted UAV Communications for Massive Multiple Access}
The number of Internet-of-things (IoT) devices is expected to surpass the 75 billion mark by 2025; thus making bandwidth resources extremely scarce \cite{Massive5G}. Hence, massive access for a sea of IoT devices 
to connect, interact and exchange data 
is becoming a daunting task. 
The ongoing urbanization 
makes this task further complicated where a large number of small IoT devices are expected to be deployed in indoor and outdoor settings \cite{rahman2019application}, and for both, shadowing and non-availability of reliable links pose a challenge. 

By integrating 
the RIS technology 
with the dynamics of the UAV, the challenges of massive access can be effectively addressed, maximizing the system capacity through optimizing the RIS phase shift vectors 
\cite{mursia2020risma}. RIS aided communications systems steer the indoor wireless channels in favor of users that demand requirements different from ordinary users. However, with RIS-assisted UAV communications systems, the viability can be extended to outdoor virtual reality (VR) applications. 
Outdoor and indoor VR users are expected to be suffer from the three major challenges: energy consumption as a result of huge data transfers,  interference from neighboring VR devices, and multi-link communications \cite{sahoo2018enabling}. 
These challenges can be effectively addressed using RIS communication systems, especially when UAVs are an integral part of them. Such a strategy is adopted in \cite{li2020reconfigurable} where the authors consider joint optimization of phase shift vectors of the RIS and the UAV altitude and trajectory to enhance coverage and capacity and enable massive connectivity. 

\subsection{RIS-assisted UAV Communications for Spectrum Sharing}
RIS can significantly reduce the interference in environments where devices are simultaneously transmitting in the same frequency band by diagonalization of the channel matrix. 
This characteristic feature makes RIS an unique choice for enabling spectrum sharing. 
Traditional spectrum sharing techniques often employ cognitive radios and demand an efficient and reliable spectrum sensing technique to minimize the interference to primary users. 
However, spectrum sensing comes at the cost of energy and reliability can be severely compromised in complex channel conditions. 

\begin{figure}[t]
    \centering
    \includegraphics[width=0.48\textwidth]{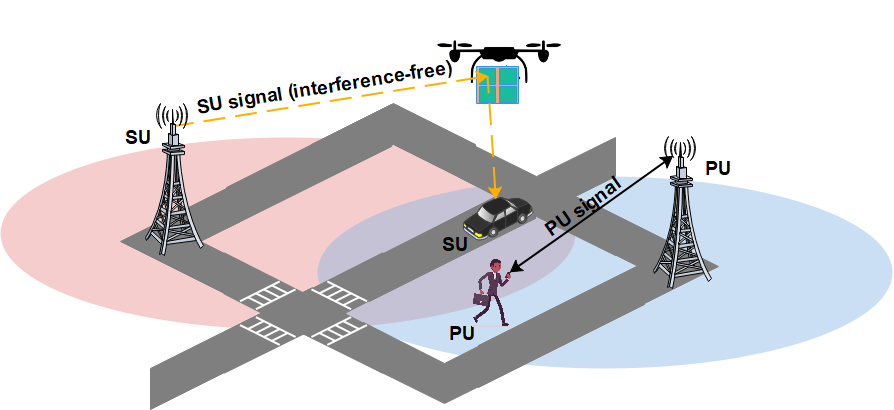}
    \caption{RIS-UAV for spectrum sharing.}
    \label{fig:Figure3}
\end{figure}

For UAV systems, the energy consumption is of high importance for long-term system operation. 
Therefore, RIS-assisted UAV systems can improve the system capacity in hot spots 
with the help of spectrum sharing. The feasibility and advantages of RIS systems supporting spectrum sharing in indoor environments have been shown empirically in \cite{tan2016increasing}, where the capacity is maximized by allowing multiple access in shared spectrum with interference among the users being controlled by optimizing the RIS phase shifters. 
In \cite{guan2020joint}, an RIS-assisted spectrum sharing strategy is proposed in order to increase secondary users (SUs) capacity while satisfying quality of service (QoS) for primary users (PUs) through phase shift optimization to achieve channel diagonalization. 
Spectrum sharing enabled by RIS-assisted UAV is a logical extension of these works, where 
the parameters describing the UAV mechanics 
will play a major role in optimizing the overall wireless networking performance in real time in dynamic user settings such scenario is introduced in Fig.~\ref{fig:Figure3}. 
Having RIS installed on UAVs, it will be important to study how altitude, longitude and latitude coordinates impact the performance of the RIS phase shifters for increasing the system capacity. 

\subsection{RIS-Assisted UAV Communications for PLS}
UAVs have been proposed for improving the PLS of terrestrial cellular networks. This is enabled by the dominant LoS links that can be established between an aerial and ground node. 
UAVs can offer different types of PLS support. For instance, UAVs can act as an AR between legitimate users to optimize the transmit power and lower the data rate for eavesdroppers. 
In addition, UAVs can be deployed as friendly jammers to send strong artificial noise (AN) that reaches 
possible attackers and protect legitimate users' privacy and data. The aforementioned roles of the UAV for improving PLS have showed great potentials in typical wireless environments~\cite{UAVSec}. On the other hand, wireless threats and attacks have been developed to create sophisticated and challenging scenarios that can degrade the performance of wireless networks even with the proposed safeguarding techniques.
An eavesdropper can, for instance, strategically deploy itself to obtain a high SNR, potentially higher than the destination node. 

\begin{figure}[t]
    \centering
    \includegraphics[width=0.48\textwidth]{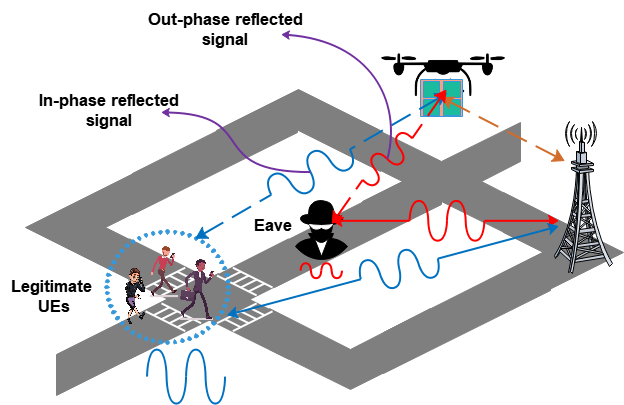}
    \caption{RIS-UAV for physical layer security.}
    \label{fig:Figure4}
\end{figure}

The use of RISs mounted on UAVs can be employed to tackle smart attackers. Prior research has shown that the secrecy rate of legitimate users increases as the distance between the communications peers 
decreases. Therefore, the free movement model of UAVs 
allows to decrease the distance between the transmission source and the desired user. After that, the phase shifts of the IRS 
can be tuned so that the reflected 
signal and the original 
signal add constructively at the legitimate user to improve the SNR. On the other hand, some of the reflecting units of IRS can employ different phase shifts 
to create a destructive reflected signal to minimize the received SNR at specific locations and limit the chances of eavesdropping. This is illustrated in Fig. \ref{fig:Figure4}.       
\subsection{RIS-Assisted UAV Communications for SWIPT}
Recently, wireless power transfer technology is being used and improved to recharge low-power devices such as IoT devices and sensors that have a limited lifetime. This can be realized through harvesting RF energy from ambient sources. The RF energy harvesting has been extended to the SWIPT concept, where the data and power can be transferred using electromagnetic (EM) waves for information decoding (ID) or energy harvesting (EH). There have been two widely used schemes for SWIPT: power splitting (PS) and time switching (TS). The PS scheme partitions the received signal in the power domain to one portion for energy harvesting and the other portion for the information transfer. TS switches between the energy harvesting and information transfer in different time slots. 

Emerging UAV technology has been proposed for assisting the SWIPT, specifically when the ID or EH devices are distributed in harsh environments or disaster areas. The reasons for this are 
the unique mobility, ease of deployment, and low cost of UAVs. As a consequence, UAVs can fly to the nearest point from the ground ID or EH devices to maximize the weighted sum rate (WSR). Also, the use of UAVs for sending the ID and EH signal will overcome the near–far problem that occurs when the ID or EH devices are distributed in a wide area with a large spacing between the ground devices. 
UAVs can act as mobile access points (APs) to 
shorten the distance between the ground devices and the AP.

\begin{figure}[t]
    \centering
    \includegraphics[width=0.50\textwidth]{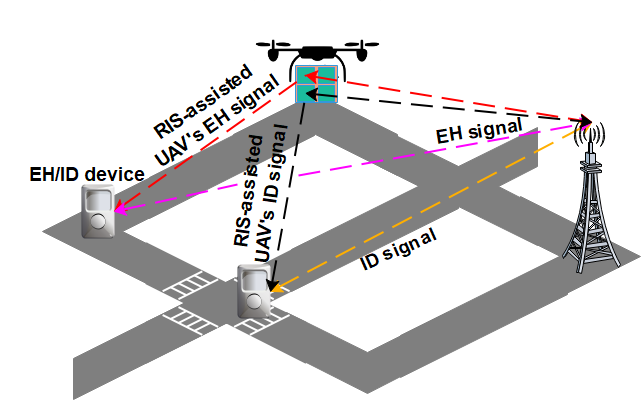}
    \caption{RIS-assisted UAV for SWIPT.}
    \label{fig:Figure5}
\end{figure}

Additionally, RIS-assisted UAV SWIPT technology will empower the performance of EH and ID for various distributions of ground devices. The deployment of RIS-assisted UAV SWIPT technology will enable additional beamforming passive beams alongside with the original EH or ID beams to maximize the WSR. 
The reflecting phase shifters 
can be adapted through the RIS controller to design the passive beamforming to optimally enhance the EH and ID performance. Fig.~\ref{fig:Figure5} illustrates this scenario where a RIS-assisted UAV is deployed 
to shorten distances and enhance the performance of SWIPT for ID and EH.
The RIS-assisted UAV SWIPT provides additional options for optimizing the performance of ID and EH. These includes the joint optimization of the 3D trajectory of the UAV and the phase shift vectors, 
\textcolor{black}{ID and EH performance and the power consumption optimization of the RIS-assisted UAV, and the optimization of the number of RIS elements and the served ID and EH users}. The planning and studying of the 3D trajectory includes defining the UAV positions for each time interval and the optimum locations where the UAV will perform the charging and information reception or transfer. The optimal design of the 3D trajectory will improve the energy efficiency of the framework by minimizing the energy consumption of the UAV enabled SWIPT and maximizing the energy transfer from the mobile UAV to the ground devices. The UAV enabled SWIPT use cases enhanced with the 3D trajectory optimization will ensure the fairness of the EH and ID among the battery-limited ground nodes and eliminate the side effects of the near-far problem. Additionally, the beamforming optimization of the UAV will focus on providing aligned energy and information beams to provide more energy charging and information transfer capacity.
The RIS-assisted UAV technology will also extend the coverage radius of the mobile charging aerial nodes to enable EH or ID in different 
scenarios, 
such as dense deployment, wide geographical distribution, and harsh signaling and disaster environments.       

Table I summarizes the key advantages of RIS-assisted UAV networks over the existing and emerging terrestrial cellular networks.




\begin{table*}[h!]
\centering
\caption{Aerial RIS-assisted networks compared to existing and emerging cellular communications technologies.} 

{\begin{tabular}{|p{1cm}|p{5.1cm}|p{5.1cm}|p{5.1cm}|}
\hline
\textbf{\footnotesize{Metric}} & \textbf{\footnotesize Terrestrial cellular networks} & \textbf{\footnotesize Integration of aerial nodes} & \textbf{\footnotesize Aerial RIS-assisted networks}
\\ \hline
\vspace{0.001 in}
\footnotesize Coverage &
\vspace{-0.05 in}
\begin{list}{\labelitemi}{\leftmargin=0.2em}
    \item {\footnotesize Different enabling technologies provide coverage extension in 5G, e.g., relaying, device-to-device (D2D), multi-hop networking, etc.}
    \item {\footnotesize However, several challenges remain and the 
    service deteriorates with channel conditions. }
    \vspace{-0.1 in}
\end{list}

&
 
\vspace{-0.05 in}
\begin{list}{\labelitemi}{\leftmargin=0.2em}
    \item {\footnotesize Aerial communications nodes can extend the coverage where and when needed. 
    } 
     \item {\footnotesize For broad coverage, a constellation of UAVs needs to be formed, which is energy consuming as well costly.}
    \vspace{-0.1 in}
\end{list}

&

\vspace{-0.05 in}
\begin{list}{\labelitemi}{\leftmargin=0.2em}
    \item {\footnotesize Aerial-RIS increases the coverage 
    by directing the beam towards the UEs 
    \textcolor{black}{in addition to 
    free positioning of the UAV 
    and ability to track and follow target UEs. 
    Additionally, a portion of the RIS elements can be used to charge the on-board battery of the UAV.}} 
    \vspace{-0.1 in}
\end{list}
\vspace{-10 in}
\\  \hline

\vspace{0.001 in}
{\footnotesize Capacity} &

\vspace{-0.05 in}
\begin{list}{\labelitemi}{\leftmargin=0.2em}

    \item {\footnotesize The driving technologies for capacity enhancements are D2D, multi-RAT communications and cell-densification.}
    \item {\footnotesize This increases the network deployment cost as well as the operational complexity and cost.}

    \vspace{-0.1 in}

\end{list}

&

\vspace{-0.05 in}
\begin{list}{\labelitemi}{\leftmargin=0.1em}
\parindent -0.2in 
    \item {\footnotesize 5G-aerial systems leverage \textcolor{black}{LoS-}MIMO to exploit channel capacity. }
    \item {\footnotesize 
    \textcolor{black}{Such can introduce severe pilot contamination with for ground users and other aerial nodes.}}
    \vspace{-0.1 in}

\end{list}
&

\vspace{-0.05 in}
\begin{list}{\labelitemi}{\leftmargin=0.2em}

    \item {\footnotesize  Aerial-RIS systems leverage \textcolor{black}{the capacity performance of terrestrial and aerial systems and boost them using the flexibility of defining the appropriate number of elements to tackle the problems such as pilot contamination. }
    Moreover, the high degrees of freedom enables increasing the spectral efficiency. 
    }
    \vspace{-0.1 in}

\end{list}

\\ \hline

\vspace{0.001 in}
\footnotesize PLS 
&

\vspace{-0.05 in}
\begin{list}{\labelitemi}{\leftmargin=0.2em}
   
    \item {\footnotesize 5G takes advantage of signal processing, e.g. precoding, to reduce the secrecy outage probability. 
    }
    \item {\footnotesize The computational complexity increases at the transmitter and the receiver. 
    }
    \vspace{-0.1 in}
\end{list}
&
 
\vspace{-0.05 in}
\begin{list}{\labelitemi}{\leftmargin=0.2em}
    \item {\footnotesize 5G aerial systems create safe zones for ground users by transmitting AN. Various communications techniques 
    can 
    prevent or mitigate jamming and spoofing.}
    \item {\footnotesize The limited energy at aerial systems may not sustain these prevention schemes. 
    }
    
    \vspace{-0.1 in}
\end{list}

& 
\vspace{-0.1 in}
\begin{list}{\labelitemi}{\leftmargin=0.2em}
    \item {\footnotesize  The aerial RIS enhances wireless security 
    by directing power to legitimate users.}
    \item{\footnotesize Since passive phase shifting is employed, significant gains in terms of energy efficiency can be achieved.}
    \vspace{-0.1 in}
\end{list}

\\ \hline
\vspace{0.001 in}
{\footnotesize Massive Access} &

\vspace{-0.05 in}
\begin{list}{\labelitemi}{\leftmargin=0.2em}

    \item {\footnotesize Different massive access techniques have been proposed, both orthogonal and non-orthogonal, 
    thus requiring sophisticated transceiver designs.}

    \vspace{-0.1 in}

\end{list}

&

\vspace{-0.05 in}
\begin{list}{\labelitemi}{\leftmargin=0.1em}
\parindent -0.2in 
    \item {\footnotesize Aerial assisted networks can be dynamically deployed closer to the end users to serve a 
    large number of devices. }
    \item {\footnotesize UAV spectrum access and interference 
    compromises scalability. }
    \vspace{-0.1 in}

\end{list}
&

\vspace{-0.05 in}
\begin{list}{\labelitemi}{\leftmargin=0.2em}

    \item {\footnotesize  An aerial RIS reflects signals to the desired users and mitigates interference. 
    }
    \item {\footnotesize Scalability can be achieved by coordinating multi-UAV-RIS systems with static RISs. 
    }
    \vspace{-0.1 in}

\end{list}
\\ \hline
\vspace{0.001 in}
{\footnotesize Spectrum Sharing} &

\vspace{-0.05 in}
\begin{list}{\labelitemi}{\leftmargin=0.2em}

    \item {\footnotesize Dynamic spectrum sharing is considered for several 5G bands for better use of spectrum.}
    \item {\footnotesize Coordination and synchronization, interference from mobile users, and MAC protocols are some of the many remaining challenges for spectrum sharing.}

    \vspace{-0.1 in}

\end{list}

&

\vspace{-0.05 in}
\begin{list}{\labelitemi}{\leftmargin=0.1em}
\parindent -0.2in 
    \item {\footnotesize Aerial assisted networks 
    provide more degrees of freedom for spectrum sharing, e.g. by limiting the interference footprint through strategic UAV positioning.}
    \vspace{-0.1 in}

\end{list}
&

\vspace{-0.05 in}
\begin{list}{\labelitemi}{\leftmargin=0.2em}

    \item {\footnotesize  Aerial RIS systems enable flexible spectrum sharing with the help of RIS, which allows multiple users to share spectrum without causing harmful interference to each other.}
    \vspace{-0.1 in}

\end{list}
\\ \hline
\vspace{0.001 in}
{\footnotesize SWIPT} &

\vspace{-0.05 in}
\begin{list}{\labelitemi}{\leftmargin=0.2em}

    \item {\footnotesize  SWIPT is suggested for terrestrial networks employing power splitting or time switching.}
    \item {\footnotesize User fairness 
    is challenging to achieve because of the near-far problem of the distributed users. 
    }

    \vspace{-0.1 in}

\end{list}

&

\vspace{-0.05 in}
\begin{list}{\labelitemi}{\leftmargin=0.1em}
\parindent -0.2in 
    \item {\footnotesize 
    The 3D mobility pattern of UAVs and LoS links can addresses the near-far problem. In addition, aerial nodes are the key enablers to provide SWIPT especially for 
    where terrestrial networks are damaged.}
    \item {\footnotesize The limited flight time of aerial nodes is a challenge, especially with dispersed users that need to be served. 
    }
    \vspace{-0.1 in}

\end{list}
&

\vspace{-0.05 in}
\begin{list}{\labelitemi}{\leftmargin=0.2em}

    \item {\footnotesize Aerial RIS-assisted systems provide 
    beams for EH and ID alongside with the 
    signals from the terrestrial or aerial network so that it can serve multiple users at the same time. The ability to continuous charge the aerial nodes by using a portion of the RIS elements 
    while the user-centered SWIPT tasks are ongoing will extend the lifetime. 
    }
    \vspace{-0.1 in}

\end{list}
\\ \hline

\end{tabular}%
}

\label{tab:survey}
\end{table*}

\section{Challenges and Research Opportunities}
\label{sec:Challenges}

\subsection{Channel Modeling}
Channel modeling is important to enable fundamental and applied research and needs 
accurate data for characterizing the path loss, as well as shadowing, scattering, and fading effects, among others. 
Various factors need to be considered for developing accurate models of RIS-assisted UAV channels. These include the ground and aerial distances, RIS fabrication material, number of elements, and the geometry of the RIS. Theoretical channel models 
will allow to investigate the ability of controlling the behavior of the communications channel and its impact in different use cases, theoretically and by means of simulations. 
The two main components of the RIS-assisted UAV channel are the UAV and the RIS that together make the channel modeling sophisticated and challenging. A UAV as an aerial node with a rapid dynamic mobility pattern and aerial shadowing modified by its movement and rotation will lead to wide spatial and temporal variations. 
The RIS adds complexity 
to defining appropriate channel models because of its passive and reflective behavior and the near-field propagation that needs to be taken into account. 
     
\subsection{Channel Estimation}
Generally, the performance of the RIS depends on optimizing the phase shift vectors as a function of the channel between the RIS and the radios. Therefore, estimating the channel between the RIS elements and the serving radios is essential and that must be determined to achieve optimal beamforming and control of the radio channel. The passive nature of the RIS elements 
characterizes this technology as  low complex and energy efficient, without the need for power amplifiers and data converters. 
However, this passive nature makes it more difficult to estimate the channel. 
Therefore, 
a variation of the passive RIS configuration is proposed 
to handle this. Specifically, the RIS elements can be accompanied with a few low-power active sensors that are responsible for sensing and estimating the radio channel. These active elements when embedded in the RIS can then be used for sending and receiving pilot signals for 
obtaining accurate channel state information (CSI). An alternative  approach 
is neglecting the estimation of the channel and relying on using machine learning algorithms, specifically reinforcement learning (RL). A Markov decision process (MDP) can be used to model the channel states 
and RL can optimize the phases in a computationally efficient manner by interacting with the channel~\cite{RLCSI}. 

\subsection{RIS Controller and Overhead}
It is important to take into consideration the controlling of the RIS elements. This controller is responsible for delivering the phase shift vectors for the antenna elements. Typically, it is assumed that the required phase shifts are transferred to the memory of the RIS controller, however, such approach demands a fully synchronized and reliable control link between the computing node and the RIS. 
This can be achieved in stationary use cases or in other cases where the control link can be easily provided. However, for RIS-UAV systems, the control link between the computing node and the RIS will face time-varying channel conditions and may experience fading and shadowing that will affect the process of uploading the phase shift modifications in real time. In addition, the size and number of elements in the RIS, which can range from a few to hundreds or more, will generate a tremendous amount of signaling overhead. 
Therefore, new solutions are needed to provide a stable control link of low latency and to limit the control signaling and processing overhead without compromising the performance. 
A possible solution for the aforementioned problem can be the use of UAV swarms that offer distributed computing and communications capabilities to ensure reliability and availability of the control links.    

\section{Emerging Technologies}
\label{sec:Emerging}

The soon to be emerging  6G wireless networks will offer a set of new technologies to meet the requirements of future wireless services and applications. Important aspects of RIS-UAVs for their integration into 6G networks is therefore discussed here. 

\subsection{Machine Learning for RIS-Assisted UAV applications}
The RIS-UAV technology will be a key enabler for delivering stunning performance for various use cases and application as discussed in Section.~\ref{sec:APPLICATIONS}. However, further  enhancements are achievable for the discussed applications when leveraging machine learning (ML) and artificial intelligence (AI) to empower intelligent RIS-assisted UAV technology. The intelligent RIS-assisted UAV will provide autonomous decision-making, knowledge extraction and prediction, and near-optimal optimization performance. The learning 
can be performed through training with labeled data, which is known as supervised learning, or without labeled data, which is known as unsupervised learning, or with real-time data, which is known as reinforcement learning. ML algorithms can be used for enhancing the RIS 
channel estimation, spectral efficiency, and balancing different tradeoffs.

The supervised ML algorithms can be deployed to train over the historical information for channel estimation, which. 
This information can be collected using conventional methods such as pilot-assisted, blind, semi-blind, and decision-directed channel estimation techniques. This labeled data 
can then be used to train the supervised algorithms to enable reliable and accurate prediction of the instantaneous channel conditions. 

However, the huge amount of data generated for training phase can be seen as a stumbling block for the limited UAV energy and flight time. Therefore, deep neural networks (DNNs) or  convolutional neural networks (CNNs) can be effectively used to extract the important features from the datasets and minimize the computational time and complexity. 
In addition, distributed ML algorithms can be adopted for reducing the computational burden on a single node; 
rather, the learning process is distributed among multiple aerial or ground nodes, or both. 

ML algorithms can 
enhance and extend the scope of current optimization techniques. 
Currently, most of optimization techniques focus on enhancing energy efficiency, data rate, and signal-to-interference plus-noise ratio. However, the emerging ML and AI will be a key enabler to find the optimum balance between different tradeoffs related to the RIS-assisted UAV performance. For example, optimizing the positioning and path planning of the UAV for a given structure and orientation of the RIS elements. Also, the on-board battery of the UAV and the weight, size, and number of RIS elements, which limit the flight time of the UAV, can be jointly designed to support a given mission. An additional tradeoff is the selection of the amount of the RIS elements to be used to reflect the incident signal 
while another part of the RIS does the energy harvesting to obtain electrical energy to charge the on-board UAV battery.  


\subsection{mm-wave and THz}
One of the unique features of future 6G wireless networks is the use of higher frequency bands, above 100 GHz. This 
will require different technology and infrastructure than used for conventional cellular networks. Mm-wave and Terahertz (THz) communications experience tough environmental conditions that result in severe attenuation and molecular absorption for high path losses. Also, the use of higher frequency band will increase the blockage rate of the transmitted signals and effect the reliability and availability of wireless communications services. 
THz communications
have shown great potentials so far in short-range LoS settings. Therefore, the proposed RIS-assisted UAV technology can become a promising solution to overcome the above mentioned obstacles. A UAV can provide the short range communication with the transmitter that uses THz bands. Also, the RIS elements are going to control the the channel conditions to provide optimum signal paths for the outgoing links. 

\subsection{VLC} 
Another emerging technology that will be used in 6G is visible light communication (VLC) because it offers low deployment costs and ultra-high data rates, and operates in unlicensed spectrum. On the other hand, there are also important drawbacks of using VLC. These are its limited coverage range, the loss of signal with every little movement or misalignment between the light emitting diodes transmitter and photo detector receiver, and the need for LoS. 
The RIS-assisted UAV framework can help overcoming these drawbacks and improving the performance of VLC. 
\textcolor{black}{The RIS for assisting VLC 
networks will have a different composition than 
a typical RIS to be able to control the incident light beams. Such RIS may use a meta-lens or crystal-liquid based RIS. The meta-lens and crystal-liquid based RIS will shape the environment of incident light signals through dynamic artificial muscles and the refractive index~\cite{VLC}.} The free mobility patterns of UAVs will enable fast and accurate alignment between the light emitting diodes and the RIS to ensure LoS. The RIS will adjust its phase shift vectors to extend coverage and increase data rates as desired. 
The RIS elements will aim to tune the shape of the light by adapting the channel geometry. This phenomena can be achieved through adjusting the RIS thickness or its refractive index~\cite{VLC}. 

\subsection{Index Modulation}
Another technology that is being actively investigated for advanced wireless communications is 
the index modulation (IM) scheme. The IM scheme relies on fetching the transmit data using indices of the available transmission entities, for example, the indices of the transmit antennas or the subcarriers of orthogonal frequency division multiplexing (OFDM) waveforms. 
The potential integration of IM schemes with the RIS-assisted communications will ensure higher spectral and energy efficiency in non-optimal SNR conditions, a lower peak-to-average power ratio (PAPR) for OFDM signals, and reduce the error rate, complexity, and cost. The RIS-based IM schemes can be deployed in different components of the system, at the transmitter, the receiver, or the RIS reflector sections~\cite{IM}. \textcolor{black}{However, the implementation of RIS-assisted IM for the transmitter will increase the control signaling overhead to allocate the activated antenna indices for performance optimization. The RIS-assisted IM schemes can be performed using a unmodulated carrier for space shift keying (SSK) or a modulated carrier for spatial modulation (SM)~\cite{IM}. The RIS-assisted UAV solution can be deployed 
close to the transmitter to ensure high spectral efficiency, specifically for the RIS-SSK scheme.
}

\section{Conclusions}
\label{sec:conclusions}
This paper provides an overview of the different services and applications that will are enabled by RIS-assisted UAV technology. We have discussed major applications of RIS-assisted UAVs and have provided a comparison with the existing and emerging technologies that extend terrestrial wireless networks. 
The  major challenges arising from integrating RIS with UAVs are presented with directions how to address them through focused research. The emerging and future technologies and the role that RIS-assisted UAVs will play beyond the presented applications in this paper are introduced as guidelines for research, implementation and experimentation. 

\section*{Acknowledgement}
A.~S.~Abdalla 
V.~Marojevic are supported in part by the by NSF SWIFT program, under grant number ECCS-2030291. 

\balance

\bibliographystyle{IEEEtran}
\bibliography{Refs,vuk}
\section*{Biographies}
\small
\noindent
\textbf{Aly Sabri Abdalla} (asa298@msstate.edu)
is a PhD candidate in the Department of Electrical and Computer Engineering at Mississippi State University, Starkville, MS, USA. His research interests are on scheduling, congestion control and wireless security for vehicular ad-hoc and UAV networks.

\vspace{0.2cm}
\noindent
\textbf{Talha Faizur Rahman} (talha.rahman@alumni.unitn.it) is a postdoctoral researcher in the Department of Electrical and Computer Engineering at Mississippi State University, Starkville, MS, USA. His research interests are on wireless communications, signal processing, and the Internet of Things.

\vspace{0.2cm}
\noindent
\textbf{Vuk Marojevic} (vuk.marojevic@msstate.edu) is an associate professor in electrical and computer engineering at Mississippi State University, Starkville, MS, USA. His research interests include resource management, vehicle-to-everything communications and wireless security with application to cellular communications, mission-critical networks, and unmanned aircraft systems.

\end{document}